\documentstyle[twocolumn,aps]{revtex}

\newcommand{\beq}{\begin{equation}}
\newcommand{\eeq}{\end{equation}}
\newcommand{\ket}[1]{\left| {#1} \right>}

\begin{document}

\title{Space, time, parallelism and noise requirements for reliable quantum 
computing} 

\author{A. M. Steane}
\address{Department of Physics, Clarendon Laboratory,
Parks Road, Oxford, OX1 3PU, England}

\date{July 1997}

\maketitle

\begin{abstract}
Quantum error correction methods use processing power to combat noise. 
The noise level which can be tolerated in a fault-tolerant method
is therefore a function of the computational resources available,
especially the size of computer and degree of parallelism. I present
an analysis of error correction with block codes, made fault-tolerant
through the use of prepared ancilla blocks. The preparation and
verification of the ancillas is described in detail. It is shown that the 
ancillas need only be verified against a small set of errors. This,
combined with previously known advantages, makes this `ancilla factory'
the best method to apply error correction, whether in concatenated or block
coding. I then consider the resources required to achieve $2 \times 10^{10}$
computational steps reliably in a computer of 2150 logical qubits, finding
that the simplest $[[n,1,d]]$ block codes can 
tolerate more noise with smaller overheads than the $7^L$-bit
concatenated code. The scaling is such that block codes remain the better
choice for all computations one is likely to contemplate.
\end{abstract}


Large-scale quantum computation can be regarded either as a realistic 
future possibility, or as a theoretical device which enables basic
questions to be asked about quantum mechanics and
the computational power of machines.
In either case, an important fundamental issue is that of the
conditions required to enable a given computational problem to be
solved on a quantum computer. Until recently, it was believed
that the effects of noise were more or less insurmountable in
a complex quantum system such as a quantum computer. However, two
discoveries have greatly shaken this belief. These are the
possibility of efficient quantum error correction for arbitrarily
large computers \cite{CS,Steane2}, and the possibility of
fault-tolerant correction and computing \cite{ft,Kita}.

An important third idea is that of concatenated coding, \cite{concat,Kitb}. 
The use of concatenated coding, combined with fault-tolerant methods, 
enables the important {\em threshold result} to be obtained 
\cite{Kitb,KLZ96,AhBO96,Got96,KLZ97,Pres97}. This result states that in 
order for arbitrarily long computations to be possible, the noise level 
$\gamma$ per elementary operation need not be arbitrarily small, but need 
only be below a threshold value $\gamma_0$. The noise must have various 
special properties for this result to be valid, such as that the effects of 
noise at different qubits and at different times are independent, that 
$\gamma$ does not grow when the computer is made larger, and that 
arbitrarily many elementary operations can be performed in parallel. 

The threshold result is an important element of the basic theory
of the control of noise in a quantum computer. However, just as it
is meaningless to talk about the amount of noise without stating
what type of noise one is dealing with, it is also meaningless to
talk about a threshold error rate without stating what other computational
resources are needed. For example, it is easy to see that the
greater degree of parallelism one assumes, the less stringent are
the requirements on memory errors in a quantum computer. The
threshold result is really a statement about how the noise requirements
scale with other requirements. The inevitable cost of the
threshold is that concatenated coding is inefficient in both space 
(number of qubits) and time (number of elementary operations), compared with 
non-concatenated codes. In view of this, it is important to compare
different coding and correction methods on the basis of all the
relevant parameters.

Fault-tolerant error correction using non-concatenated codes is very 
powerful if we take advantage of the approach I put forward in an earlier 
paper \cite{Steane97}. In this approach, the task of error correction is 
reduced essentially to one of quantum state synthesis. The most important 
task for a quantum computer is to generate verified copies of the encoded 
zero state $\ket{0_E}$ in ancilliary blocks of qubits. I will refer to this 
method as the `ancilla factory' method. 

For sufficiently long computations, a concatenated correction method is 
probably the best choice because of the threshold result. However, the 
estimates in \cite{Steane97} suggest that the size of computation (number of 
logical qubits times number of computation steps) may well be very large ($> 
10^{15}$) before concatenated coding is preferable to simple block coding 
with the ancilla factory method. The present paper will elucidate this by 
providing a more complete analysis of the ancilla factory than was done 
previously. In particular, the network required to verify the ancilla state 
is fully specified and taken into account. It is noteworthy
that this network performs a minimal error detection, in a sense
to be described. Also, the usefulness of measuring the syndrome
and using classical processing to interpret it, rather than a further
quantum network, will be underlined.
I will show that the ancilla factory method makes error correction using CSS 
(Calderbank, Shor and Steane \cite{CS,Steane2}, see also \cite{Got97}) codes 
sufficiently efficient that it compares favourably with the concatenation 
method, even for a large computational task such as factorisation of a 130 
digit (430 bit) number using Shor's algorithm. 

The main conclusion of this study is a `bench mark' indication of the power 
of different error correction methods, in terms of four parameters: the 
noise level, the `scale up' (increased number of qubits), the `slow down' 
(increased number of operations) and the degree of parallelism (ability to 
apply many gates simultaneously) required to make a computer sufficiently 
reliable to carry out a specific difficult task. This bench mark is offered 
as a target for others to bring down, either by optimisation of current 
methods or by finding radically new ones. It is a rough guide to the 
hardware which would be needed if large scale quantum computations were to 
be attempted one day.

\section{Choice of method}

There are two requirements for error correction techniques
to be used to stabilize a quantum computer. These are fault-tolerant
error correction, and fault-tolerant computational operations.
Two approaches to fault-tolerant error correction were put forward
by Shor \cite{ft} and Kitaev \cite{Kita}. In Shor's approach, the
quantum error correcting code (QECC) is taken from a subset of the CSS
codes described in \cite{CS,Steane2}, and the syndrome extraction is carefully
arranged so as to restrict the propagation of errors. In Kitaev's
approach the QECC is itself designed so that each qubit is
involved in a bounded number of check operators (a `local check code'),
so that even a straightforward extraction of the syndrome will not
propagate errors too far. Shor's fault-tolerant syndrome extraction
was generalised to all stabilizer codes \cite{Gott,CRSS}
by DiVincenzo and Shor \cite{DiVS}. The ancilla factory 
method which I have proposed is closely related to Shor's ideas.
Since it is to my knowledge the best method currently known, it is used 
in this study. 

Aharanov and Ben-Or \cite{AhBO96} have discussed fault-tolerant methods
in which syndromes are not measured (ie converted into classical
information) during the computation, but are interpreted within
the unitary quantum evolution. The advantage of avoiding measurements
is that the analysis is simpler, and measurements will almost
certainly be slow compared to unitary operations in an experimental
system. However, the disadvantage is that the quantum computer
must be made much larger in order to store the extra `garbage'
manufactured by error corrections, and measurements provide
robust classical information which is very useful since much of
the syndrome interpretation can then be done classically.
For these reasons I retain the use of measurements.

Fault-tolerant computational operations have been discussed for various types 
of stabilizer code. A universal set of operations was described by Shor 
\cite{ft} 
for CSS codes obtained by puncturing a self-dual classical code whose words all 
have weight equal to a multiple of four (`doubly even' codes). Further methods 
were described by Knill {\em et. al.} \cite{KLZ96}
for codes obtained by puncturing one with 
with weights divisible by 8 and dual distance $\ge 4$. The same authors
give some more techniques in \cite{KLZ97}, based partly on the ideas
of Kitaev \cite{Kitb}. Rotations, such as those needed in a
Fourier transform network, are considered by Preskill \cite{Pres97}.
Gottesman has described methods
by which a universal set of operations can be obtained for any
stabilizer code \cite{Got97}, though some of these methods are considerably
more complicated than the ones designed for particular codes.

The general rule is that the
more symmetry and simplicity there is in the code construction,
the simpler are the fault-tolerant methods. This study will be restricted
to quite simple codes, namely $[[n,1,d]]$ CSS codes obtained
by puncturing a doubly even classical self-dual code. The notation $[[n,k,d]]$ 
refers to a QECC encoding $k$ qubits into $n$, with minimum distance $d$, thus 
$t=(d-1)/2$-error correcting. Note that we choose to encode only a single 
qubit in each block, and restrict ourselves to doubly-even codes
in order to allow a large set of bitwise operations.
It may well be worth while to sacrifice some of the 
simplicity in the computational operations in order to gain more efficient 
storage; this is an avenue for future work.

\section{Assumptions}

In order to estimate the reliability of the encoded quantum computer,
we will make a set of standard assumptions about the noise, which have been
discussed by several authors \cite{concat,KLZ97,Pres97}. Note that much more 
general error models have been considered by Knill {\em et. al.} 
\cite{Knill,KLZ96,KLZ97}. The noise model assumes independent
stochastic errors among 
the qubits, whose overall effect can be estimated by adding classical error 
probabilities rather than quantum amplitudes. In other words, to keep the 
problem managable we will only try to find out the limitations imposed by such 
random errors. 

We will adopt the depolarising channel model for the noise on each qubit 
\cite{Benn}. For a given error opportunity, a qubit remains 
unchanged with probability $1-p$, and undergoes $\sigma_x$, $\sigma_y$ 
or $\sigma_z$ errors with equal probabilties $p/3$. Of all the errors 
that occur, on average $2/3$ will require correction in the computational basis 
$\{\ket{0},\; \ket{1}\}$ and $2/3$ in the conjugate basis $\{ 
\ket{0}+\ket{1},\;\ket{0}-\ket{1} \}$. The errors $\sigma_x$ and $\sigma_z$
will be referred to as `bit errors' and `sign errors' respectively.

To model the noise in a gate operating on two qubits, we assume the gate 
produces the action it should, followed by error operators. 
There are sixteen possible error operators,
all the members of the set $\{I, \sigma_x, \sigma_y, \sigma_z\}$
$\otimes$ $\{I, \sigma_x, \sigma_y, \sigma_z\}$. If the gate error
probability is $p$, we assume the error-free case $I \otimes I$ occurs
with probability $1-p$, and the other 15 cases occur with equal
probabilities $p/15$. However, we will not take advantage of the fact
that errors in the two qubits are correlated (for example if a bit-error is
found in the first qubit, the probability of bit-error in the second
qubit rises from $(8/15)p \ll 1$ to $1/2$.) Rather, we assume that with 
probability $O(p)$ (not $p^2$) errors are produced in both qubits, but we 
must correct the errors as if they were independent. Thus we over-estimate 
the difficulty of error correction. Furthermore, we will overestimate the 
probability $8/15$ as $2/3$, to simplify the equations. 

As usual in quantum error correction, when we talk of errors occuring
with given probabilities, we refer to the probabilities that the quantum
state is projected onto each state of well-defined
error syndrome, when the syndrome is later extracted. We allow
for relaxation in the system by permitting
states of different syndrome to be entangled with different environment
states \cite{Steane2,Knill}.

The most important, and possibly unphysical, assumption is that
error probabilities can be multiplied. That is, if one error
has probability $p_1$, and another $p_2$, the probability of
getting both is $p_1 p_2$. This is reasonable if errors at different
places and times are uncorrelated with one another, and it corresponds
to the standard Gaussian noise model commonly used in physics. However, noise 
in real systems often has two components: one which can be well described in 
terms of standard Gaussian statistics, where the probability of obtaining an 
event many standard deviations from the mean falls exponentially with the 
distance from the mean, and another component in which the wings of the 
probability distribution are much higher. In other words, rare events are much 
more likely in practice than one would suppose on the basis of the standard 
treatment of experimental error. The correction methods discussed here 
cannot provide stability against such noise. In order to handle it,
others methods must be used in conjuction with the error correction
we are considering.

The overhead required to implement error correction depends on the
noise level. When estimating the scaling of this overhead with the
noise level, it is convenient to assume that the noise level per
qubit is independent of the total number of qubits in the computer. 
Such an assumption is required, for example, for the threshold result for 
concatenated codes. This is unlikely to be valid for a real device. However, 
when giving the noise level required for a task on a computer with a specified 
number of qubits, no assumption about the scaling of noise need be made.

Another consideration is that of parallelism, ie the number of
quantum gates that can be executed in parallel in a single time step.
A large degree of parallelism turns out to be necessary 
in order to keep storage errors under control. For this reason
we will be explicit about the degree of parallelism involved in the
methods to be discussed.

Most of the quantum gates in the stabilized computer will be
involved in error correction rather than the execution of a computational
step. For this reason most two-qubit gates only connect qubits within
a few blocks of one another. However, a computational step might
involve pairs of qubits anywhere in the computer. It will be assumed
that such gates can be applied to any pair of qubits without any significant
extra cost.

Fault tolerant quantum computation makes much use of state preparation
and measurement. These irreversible operations involve an
amplification of some kind\footnote{Roughly speaking, this is an amplification
from the quantum to the classical level.}
if they are to be done reliably.
They are therefore much slower than a unitary quantum gate. 
In a practical system one would probably wish to allow many
gates to be carried out on other (unrelated) qubits during the time any given
qubit is prepared or measured, but this will complicate
the analysis here. We will allow the same time (in fact one time step)
for a gate as for a measurement. Therefore the results are strictly
valid only if the time step is taken to be the time necessary for
a measurement rather than a gate. However, there are
many more gates than measurement operations in the networks actually
used, therefore it is a fair approximation to take the time step
to be that of a gate. 

A further question of speed of operation is that involved in decoding
of the error syndromes. This is done by a dedicated classical computer,
and we assume that the result is available within the time required
to prepare an ancilla state in the quantum computer. This is
itself quite a large classical computation, involving decoding
thousands of $(n-1)$-bit syndromes, where $n=23,55,87$ for the
codes to be discussed.

Finally, the important question of leakage errors is here assumed to
be solved, for example by converting leaked qubits into erroneous ones,
using leak detection circuits as in \cite{Pres97}. The overhead
represented by such additional circuits will not be included in 
the analysis.

\section{Analysis}

We will adopt the following notation. The computation to be
performed requires $K$ logical qubits and $Q$ computational steps.
This is achieved in a set of $N > K$ physical qubits using
$T > Q$ elementary gates. $N$ is of order $n K$, where $n$ is 
the block size of the $[[n,1,d]]$ CSS code used. The number
of correctable qubit errors per block is $t = (d-1)/2$.
The parameters $K,Q,N,T,n,t$ are integers. The noise is
parametrised by two real numbers, 
$\gamma$ the gate or measurement error probability, and
$\epsilon$ the store error probability per qubit per time step.

The parameters of the logical computer are $K$ and $Q$.
This logical computer is embedded in a physical computer of 
parameters $N$ and $T$. We use the language `computational
step' to mean a gate on encoded qubits which evolves the logical
computation, and `elementary gate' for a gate on bare qubits.
The main overheads involved
in error correction are the `scale-up' $N/K$ and the `slow-down'
$T/Q$ (number of gates per computational step). The amount of
time required to finish the computation is of order $T$ divided by the
degree of parallelism.

We will analyse a complete fault-tolerant computation, using
correction by a block code, and then compare with the
requirements for a concatenated code, drawing
on the work of Preskill \cite{Pres97}.

The `ancilla factory' error correction method is based on the generation
of the encoded zero state $\ket{0_E}$, as described in \cite{Steane97}.
The $n$-qubit block to be corrected will be referred to
as $b$, and the $n$-qubit ancilla as $a_x$. A further qubit
is used for verification of $a_x$. We imagine that error
correction is carried out in parallel on all the $K$ blocks
in the computer. However, within a block we will restrict parallel
operation to single-bit gates\footnote{If this assumption is relaxed,
as for example in \cite{Pres97}, the restriction on memory error rate
is less severe.}.
The state $\ket{0_E}$ is generated in $a_x$
and then rotated by the Hadamard transform $H$ applied bitwise.
Since we are using $[[n,1,d]]$ CSS codes this produces 
the state $\ket{0_E} + \ket{1_E}$ (ignoring normalisation). The network
to perform these operations is given in fig. 1 for the 7-bit code $[[7,1,3]]$ 
\cite{Steane1}. The general features of the method were described in 
\cite{Steane97}, but here we will be specific about the networks and the 
propagation of errors. The network in fig. 1 is designed both to generate and 
to verify the state required. If an error is detected, the network is
restarted from the beginning.
The verification only picks up bit errors (Pauli 
$\sigma_x$ operator) in the state, not sign errors (Pauli $\sigma_z$ operator). 
This is because bit errors would be converted into sign errors by the Hadamard 
transform, and then carried into $b$ by the subsequent {\sc xor} operations, so 
we must guard against them. Sign errors will be converted into bit errors by 
the Hadamard transform, and result in an incorrect syndrome. The fact that sign 
errors (before the $H$ gates) can propagate from one qubit to many is not a 
problem since in any case we will assume that a single sign error makes the 
syndrome invalid. A bit error, before the verification gates, can also 
propagate from one qubit to many, but such errors will be detected and the 
ancilla preparation restarted. The remaining undetected bit errors are
uncorrelated.

Meanwhile a separate ancilla $a_z$ is prepared, in parallel (see fig. 3), to 
be used for correction of sign errors in $b$. The preparation network is the 
equivalent of fig. 1 but in the conjugate basis. 

Let us first confirm the claim about the verification of the ancilla, namely
that only uncorrelated sign (bit) errors are finally coupled from
$a_x$ ($a_z$) into $b$. The network of
fig. 1 can be divided into three parts: generation, verification, and
Hadamard transformation.
First assume the verifier gates are noise-free. Then the most
general state which can survive the verification is
  \beq
\sum_{i \in C} c_i \ket{i} \otimes \ket{e_i}   \label{verif}
  \eeq
where $C$ is the classical $[7,3,4]$ code and $\ket{e_i}$ is
a state of the environment. 
In order to analyse the coupling to $b$
we make use of the simple equivalence shown in fig. 2.
The most general error-free state of $b$ is 
$\alpha \ket{0_E} + \beta \ket{1_E}$. 
The Hadamard transformation of $b$ (see fig. 2) carries this state to
$\sum_{j \in C} (\alpha + \beta) \ket{j} + (\alpha - \beta) \ket{\neg j}$,
written using the same basis states as eq. (\ref{verif}). 
To obtain this form 
we have made use of the CSS code construction $C \in C^{\perp}$.
The {\sc xor} between $a_x$ and $b$ produces
  \begin{eqnarray*}
&&\sum_{i \in C} c_i \ket{i}\ket{e_i} \sum_{j \in C}
(\alpha + \beta) \ket{j+i} + (\alpha - \beta) \ket{\neg j +i} \\
&=& \sum_{i \in C} c_i \ket{i}\ket{e_i} \sum_{j \in C} 
(\alpha + \beta) \ket{j} + (\alpha - \beta) \ket{\neg j}
  \end{eqnarray*}
since displacing a linear code by a member of the code simply
moves the code onto itself. Hence the most general noisy state of $a_x$
which survives noise-free verification produces no errors in $b$.
The significant point is that this is achieved with only a 
weak verification: one which detects none of the
sign errors and only those bit-error vectors which are not in $C$, which
is a small set (only $2^4$ out of $2^7$). 

When we allow the verification to be itself noisy, its gates introduce
both bit and sign errors. However, the verification network
is one in which bit errors can not propagate among the qubits of
$a_x$, therefore the bit errors which are produced are uncorrelated.
The same is true for the Hadamard transformation of $a_x$ at the end
of fig.~1. 

The network of fig. 1 involves 46 elementary operations and 30 time
steps. An elementary operation is taken to be a gate applied to bare (non 
encoded) qubits, or a preparation or measurement of a bare qubit. It is assumed 
that single-qubit elementary operations can be carried out in parallel with 
double-qubit ones. More generally, if the generator matrix of the classical
code giving $\ket{0_E}$ in the quantum code has $m$ rows, and the average
weight of a row is $w$, then the numbers of operations and time steps
are as follows.

There are $n+1$ preparation operations, $m+1$ measurement operations,
and $n+m$ Hadamard gates, making $2(n+m+1)$ single-bit operations. 
There are $m(w-1)$ {\sc xor} gates between qubits in $a_x$,
and $m(w+1)$ {\sc xor} gates between $a_x$ and the verification bit,
where we have used
the fact that the code is $[[n,1,d]]$, so the final parity check in
the verifier needs $m$ gates (this verifies that $\ket{0_E}$, not
$\ket{1_E}$, has been generated). The number of time steps is
  \beq
m(w-1)+2+m(w+1)+m+1 = m(2w+1)+3.    \label{ta}
  \eeq
For the $[[n,1,d]]$ CSS codes which we are considering, $m=(n-1)/2$ and $w 
= d+1$ \cite{Steane2}. The number of time steps is thus $nw + (n+5)/2 - w$. 
This completes the preparation of $a_x$. 

Once the ancilla state is prepared, $a_x$ and $b$ interact by an {\sc xor}
gate applied transversally, ie $n$ elementary operations, and then
the ancilla is measured. These {\sc xor} gates are applied sequentially,
not in parallel. The syndrome is extracted from the measurement
results. Let $1-\alpha$ be the probability that the correct syndrome is 
obtained. This requires no bit errors to be present in $a_x$ when the 
syndrome is extracted, whether from gates or free evolution, so 
  \beq
1 - \alpha \simeq \left( 1 - \frac{2\gamma}{3} \right)^{2(mw+2n+m+1)}
\left( 1 - \frac{\epsilon}{3} \right)^{(n-1)(m(2w+1)+3)}.
  \eeq
In writing this expression, we have taken into account the fact that most 
$\sigma_y$ errors produced in $a_x$ before the final verification will be 
detected, so we only need account for $\sigma_z$ memory errors, of 
probability $\epsilon/3$. This is also why the error probability associated 
with the {\sc xor} gates is accounted $2 \gamma/3$ rather than somewhat 
larger. Two of the timesteps (initial preparation, and the Hadamard 
transform) do not involve any free evolution. Apart from this, we have 
neglected the slight reduction in storage errors obtained by preparing 
qubits just before they are needed. We will find later that we require 
$(\alpha/n)^{t+1} \ll 1/nKQ$, so $\alpha$ is small (eg $\sim 0.001$) and 
most of the time the ancilla preparation is succesful. The ancilla 
verification plays an important role in rejecting failures, but these are 
rare occurences. 

The ancilla preparation and syndrome extraction is repeated at least $r$ 
times before any attempt is made to correct the block $b$. An error may 
well be generated in $b$ during this process, in which case the syndrome 
should change. Therefore in making the best guess of what corrective 
measure to apply to $b$ on the basis of the syndrome information obtained, 
allowance must be made for such possibilities. In this analysis I will 
assume that on average $r = t+1 = (d+1)/2$ syndromes are needed. The logic 
of this choice will become apparent shortly. Usually the first $r$ 
syndromes will be consistent (probability $\sim (1-\alpha)^r > 0.99$). 
Either they agree, or they are consistent with an identifiable error 
developing during the free evolution of $b$. When they are not consistent, 
usually just one more syndrome will suffice to produce an unambiguous 
interpretation, though we allow ourselves to generate $O(r)$ more 
syndromes, before correcting, on the rare occasions that this is necessary. 

The probability of getting a set of $r$ syndromes consistent with one 
another but all incorrect, and thus of applying the wrong `corrective' 
measure to $b$, can be estimated as the probability that $r$ successive 
ancillas all develop an error in the same qubit\footnote{This part of the 
analysis differs from \cite{Steane97} which was overly cautious at this 
point.}. This is of order $2 n (\alpha/n)^r$. We consider the whole 
computation to fail if this happens. We will find shortly that a much more 
likely cause of failure is that too many errors accumulate in $b$, so even 
if we have greatly underestimated the probability of using the wrong 
syndrome, the final conclusions about error rates will be unaffected. The 
choice $r=t+1$ is perhaps an `overkill', but it is the logical choice in 
order to ensure that the syndrome failure probability scales with $t$ in 
the same way as the probability of generating too many errors in $b$. 

We now wish to find how many errors accumulate in $b$ during the course of 
the computational steps, the ancilla preparations, and the interaction 
between $a$ and $b$. It is very important to allow for errors in the free 
evolution of the qubits (storage errors) as well as gate errors, since the 
qubits spend much more time `resting' than in gate operations. 

Most of the operations on any block $b$ come from the error corrections 
rather than the computational steps. This is because most computational 
steps involve operations on only a few of the $K$ logical bits. The 
remaining $K-2$ or so blocks freely evolve and then require correction for 
storage errors, and for the gate errors introduced by the last correction. 
Occasionally, however, single-qubit gates are applied to many, even all of 
the $K$ logical bits, so we will allow for the errors caused by one 
transversal operation on $b$, representing one computational step, every 
time we apply error correction to $b$. That is not to say that on average 
we correct the whole computer once for each step in the computation. 
Rather, we can fit in many computational steps, as long as they act on 
different blocks, between corrections. Let $\eta$ be the average number of 
computational steps per correction of the whole computer. Since there are 
more than $n w$ time steps needed to prepare the ancillas (eq. (\ref{ta})), 
a safe choice is $\eta = w$, where we suppose the computational steps are 
carried out during the first of the $r$ ancilla preparations. This is 
consistent with our assumption that operations on different blocks (here, 
$a$ and $b$) can be done in parallel, and it keeps computational and 
correction operations on $b$ completely seperated in time. The optimal 
choice for $\eta$ may be considerably larger than $w$, so we are being 
cautious. 

We will first count the bit errors in $b$. We then assume the same number 
of sign errors arise, and both correctors (bit and sign) must succeed 
if the computation is not to fail. 

The {\sc xor} interaction between $b$ and the first ancilla, $a_x$,
involves $n$ gates for each repetition of 
the syndrome generation, making $rn$ gate error opportunities, each with
probability $\sim (2/3)\gamma$ to leave a bit error in $b$.
The {\sc xor} 
interaction between $b$ and the second ancilla $a_z$ provides a further $rn$ 
opportunities and also carries bit errors from $a_z$ into $b$. The gate 
errors which escaped detection in the verification of $b$ are
mainly those 
occuring in the last verification ({\sc xor}) gate applied to each qubit, 
and in the $H$ gates, making $2rn$ opportunities. Note that 
these error opportunities are uncorrelated with each other and with
the other noise in $b$. The single computational step 
affecting any given block $b$ before correction produces $n$ gate error 
opportunities. The total number of independent
gate error opportunities which the 
$\sigma_x$ correction must deal with is therefore 
  \beq
g = n(4 r + 1).  \label{g}
  \eeq

Each ancilla preparation involves $nw + (n+5)/2 -w$ time steps (eq. 
(\ref{ta})). A further $2n+1$ are needed for the {\sc xor} gates and 
measurement of the two ancillas. With $n$ qubits in $b$, this gives
$n(nw + (n+5)/2-w +2n+1)$
opportunites for storage error in $b$ for each cycle of 
syndrome generation. In addition, there are approximately $n^2/2$ 
opportunities for undetected storage errors which are carried from $a_z$ into 
$b$. The total number of storage error opportunities is 
  \beq
s_s = n\left((n-1)w + \frac{5}{2}n + \frac{7}{2}\right) r  \label{ss}
  \eeq
when the $r$ syndromes are generated in series (fig. 3a), and
  \beq
s_p = n\left((n-1)w + \frac{1}{2}n + \frac{7}{2} + 2nr \right) \label{sp}
  \eeq
when they are generated in parallel (fig. 3b). 

The probability that either more bit-errors, or more
sign-errors, or both, accumulate in $b$ than can be
corrected is
  \beq
P \simeq 2 \sum_{i=t+1}^g \frac{g!}{i!(g-i)!}
\left( \frac{2}{3}\gamma + \frac{s}{g}\frac{2}{3} \epsilon \right)^i  \label{P}
  \eeq
where $s$ is either $s_s$ or $s_p$, and
we used the fact that for $\epsilon \ll 1$, it is equally likely
for something
of probability $\epsilon$ to occur $s/g$ different ways as it is for
something of probability $\epsilon s/g$ to occur once. Recall
that $w = 2t+2$ and $r=t+1$ so $s_s/g \simeq nt/2$.

\section{Code comparison}

The value of $P$ (eq. (\ref{P})) is plotted in fig. 4 as a function
of $\gamma$, for various CSS codes with the required properties.
In fig. 4 (a) serial syndrome repetition is assumed, and
the case $n \epsilon = \gamma / 2$ is plotted. Fig. 4 (b) shows
results using parallel syndrome repetition, taking
$n \epsilon = 2 \gamma$. It is seen that 
parallel repetition allows the memory noise to rise by a factor
$\sim 4$, for given $P, \gamma$, at the cost
of increased storage and parallelism overheads.
In all the cases plotted, storage errors dominate $P$.
The codes are $[[7,1,3]]$, 
$[[23,1,7]]$ obtained from the famous Golay code
\cite{Golay}, and $[[55,1,11]]$ and $[[87,1,15]]$ obtained from
self-dual double circulant codes with weights divisible by 4
\cite{MacWS}. Note that slightly more efficient (quadratic residue)
CSS codes exist \cite{SQECC}.

$P$ is the failure probability per qubit per correction. To
achieve a computation of $Q$ steps involving $K$ logical qubits,
we require $P < \eta / KQ$. The overhead involved in the
ancilla factory method, applied to the block codes we have
chosen, is
  \begin{eqnarray}
\frac{N}{K} &=& \left\{ \begin{array}{ll}
n + 2(n+1) & \mbox{2 parallel ancillas per block} \\
n + 2r(n+1) & \mbox{$2r$ parallel ancillas per block}
\end{array} \right., \\
\frac{T}{Q} &\simeq& \left( (n-1)w + 5n \right) 2r K / \eta.
  \end{eqnarray}
The degree of parallelism needed is $3 K$ or $K(1+2r)$
using serial or parallel syndrome accumulation.

For comparison, a concatenated code based on the 7-bit
code has overhead
  \begin{eqnarray}
\frac{N}{K} &\simeq& \left( \frac{ \log \gamma_0 Q }
{ \log \gamma_0 / \gamma } \right) ^{\log_2 7}, \\
\frac{T}{Q} &\simeq& 7^{L-1} \, 480 K / \eta
  \end{eqnarray}
where the number of levels $L \simeq \log (N/K) / \log 7$. The expression 
for $N/K$ is from \cite{Pres97}, and I estimate $T/Q$ by noting that $(1 + 
1/2 + 1/4 + \ldots) 7^{L-1} Q/\eta$ $7$-bit recoveries must be carried per 
logical qubit, if 2 recoveries are carried out per level. Each correction 
uses two ancillas as in fig. 1, and the syndrome is generated twice, making 
$4 \times 60$ elementary operations\footnote{Using Shor's `cat state'
method to evaluate the syndrome, the factor 480 becomes 576.}. The degree 
of parallelism required is approximately $2 K 7^{L-1}$. 

To compare methods, let us consider the case suggested by Preskill 
\cite{Pres97}, in order that we can use his estimates for concatenated coding. 
The example case is that of factorising a 130 digit (430 bit)
number using Shor's algorithm. This factorisation problem would take
a few months on current classical machines with the best known classical
algorithm. Shor's algorithm requires 
$K \simeq 5 \times 430 = 2150$ encoded qubits, and of order
$10^9$ Toffoli gates, using the circuit of Beckman {\em et. al.} \cite{Beck}.
Using Shor's fault-tolerant method \cite{ft,Pres97}, each Toffoli
gate requires about 20 fault-tolerant operations such as {\sc xor}
or state preparation. Therefore we will take $Q=2 \times 10^{10}$, and
the ancilla factory with $[[n,1,d]]$ block coding 
will succeed if $P < 2 \times 10^{-14} \eta$. Tables 1 and 2 show the 
error rates and overheads needed to achieve this for various CSS codes.
For comparison, the more complete analysis reported by Preskill
for the the 7-bit code with
three levels of concatenation gives $\gamma \simeq \epsilon \simeq 10^{-6}$,
with $N \simeq 1.3 \times 10^6$ qubits \cite{Pres97}.
I estimate $T \sim 10^{17}$ and a required parallelism of
order $2 \times 10^5$ for this case.

Table 1 gives the results when the $r$ syndromes are extracted in series
(eq. (\ref{ss})), and table 2 gives the results when the $r$ syndromes
are extracted in parallel (eq. (\ref{sp})). Examining table 1, we find
that the ancilla factory
allows the gate error rate with $[[55,1,11]]$ block coding to be 
an order of magnitude larger than the result for concatenated coding quoted 
by Preskill, but requires the storage error rate to be an order of 
magnitude smaller. Since gate errors are liable to be more noisy
than free evolution in practice, the block code here compares favourably
with the concatenated code.
Since the block coding method also has a smaller overhead, the
conclusion is that it is the best method in this example.

Examining table 2, we find that block coding with parallel
syndrome repetition achieves roughly an order of magnitude
improvement in the allowed gate error rate, without much change
in the store error rate, compared to the concatenated code,
with a similar size overhead but smaller time and parallelism.
Here again it is the better choice.

\section{Discussion}  \label{s:d}

It should be underlined that the comparison in tables 1 and 2
only gives a rough indication of the relative merits of concatenated
and block coding, because the analysis I have used is approximate,
whereas the results quoted in \cite{Pres97} are based on a
more careful computer simulation of the flow of errors through
the relevant quantum networks. The present results show that
a more thorough comparision would be worthwhile. It would
be surprising if such an analysis overturned the overall conclusion,
however, since that would require the results here presented to
be incorrect by about an order of magnitude. 
Since my result is essentially $P \sim O((g \gamma)^{t+1})$ this
would require my estimate of $g$ (eq. (\ref{g}))
to be incorrect by an order
of magnitude. However, I have been cautious at various points
in the calculation, especially in supposing 
the syndrome repetition $r$
to be on average as large as $t+1$, so I expect my results
to be cautious rather than over-optimisitic. Zalka \cite{Zalka}
has suggested accepting the first syndrome, without repetition,
whenever it indicates no errors. Such a method would allow
$g$ to be reduced, and therefore $\gamma$ to be increased, by
about a factor $t$. Also, it
would probably be safe to increase $\eta$, the average
number of computational steps per correction. Such simple adjustments would 
probably allow the error rates given in the tables
to be improved by about an order of magnitude. 
Supporting evidence for this statement is obtained
by observing that the threshold $\gamma_0$ for $7^L$-bit concatenated
coding in the case of no memory errors ($\epsilon =0$) is about
an order of magnitude larger in Zalka's optimised method \cite{Zalka}
than in the corresponding results of
Gottesman {\em et. al.} \cite{Got96,Pres97}.

Let us now estimate the size of computation for which the $7^L$ concatenated 
code would be preferable to block coding. With concatenation, when we require 
more computational steps the overheads $N/K$ and $T/Q$ increase, but the 
required noise level $\gamma, \epsilon$ is roughly unchanged. Let us add one 
more level of encoding, $L=4$, and compare with the $[[87,1,15]]$ used with 
parallel ancillas (table 2). The noise requirements of the two methods become 
roughly equivalent if we allow $\gamma$ and $\epsilon$ for the $[[87,1,15]]$ 
code to be reduced by a factor of 3 from the values in table 2. The block 
code then requires 7 times less memory noise, but allows 6 times more gate 
noise, than the values $10^{-6}$ for the concatenated code. Since the block 
code requires only a third as many qubits, and much smaller time and 
parallelism, one may argue that it is still the better choice. In any
case, the value of $P$ will fall by $3^8 = 6561$, so the computation
size can increase by the same factor. This would permit Shor's algorithm
on numbers of 9 times more digits, ie a 1170 digit (3890 bit) number.
The important point is not in the details, but simply that block coding 
is the better choice for all the computation we are likely to contemplate. 

Another consideration is the relative ease
of applying computational steps fault-tolerantly with
the various codes we have considered. The 7-bit code
has the advantage of being simple, and so fault-tolerant
operations are relatively easy to find. However, after two
or three levels of concatenation some of the fault-tolerant
operations become quite complex, especially those requiring
preparation of a given state (in encoded form). 
Block coding is advantageous in this respect.
The $[[23,1,7]]$
code is obtained from the famous Golay code \cite{Golay} which
is certainly a good choice for fault-tolerant operations because
of the many symmetries of the extended Golay code $[24,12,8]$.
(for example its weight distribution is
$A_{\{0,8,12,16,24\} } = \{ 1, 759, 2576, 759, 1 \}$). 
An interesting possibility is to concatenate the Golay code
with the 7-bit code. This would produce a similar correction
ability to the $7^3$ concatenated code, but with half the block
size. This may well allow the `benchmark' of tables 1 and 2 to
be improved upon.

An important lesson is that it is not necessary to consider
infinite families of codes (such as the Reed Muller codes)
when considering the practical possibilities of quantum computation,
since the finite set of {\em known} codes already contains specimens
sufficiently powerful for all the quantum computation we are likely
to contemplate.

\section{Ancilla factory}

Entropy is kept under control in quantum computation by the
feeding in of low-entropy prepared states. At the most
basic level, one can take these low-entropy prepared states to
be states of bare qubits. However, Shor's fault-tolerant
method made a significant step forward in focussing attention
on the low-entropy preparation of entangled states (the so-called
`cat' states). The ancilla factory pushes this a stage further.
One may wonder whether there is more to be gained by taking another
step in the same direction. This would involve preparing
well-chosen entangled states of larger ancillas, aiming to
reduce the number of ways in which errors can be produced
in, or coupled into, each block $b$ during correction. However, 
unless some radically different approach to quantum
error correction can be found, there is little further to
be gained in this direction. The reason is that any corrector
must gather sufficient information from $b$ to learn
the bit and sign errors in all the $n$ bits of $b$.
In order to couple this information out of $b$ into some
other system (whether quantum or classical), at least $2n$
{\sc xor} gates must operate on $b$.
If on average the number
of syndrome repetitions can be $r \simeq 1$ rather than
$t+1$, which seems likely, then this minimum is achieved
by the ancilla factory for all stabilizer codes \cite{Steane97}.

The main possibility for further improvements is in the preparation
of the ancilla state. Fig. 5 shows a network to prepare a
7-bit ancilla for the $[[7,1,3]]$ code using fewer gates than 
fig. 1. The idea is to take advantage of the structure in the
generator, in which not all error propagation routes are open,
so that the verifier need not carry out all the
parity checks in the $[7,3,4]$ check matrix. Similar networks
exist for more complicated codes.

In this paper I have supposed that the only way the quantum computer can be 
manipulated is through quantum network methods, ie preparation 
and measurement of qubits in 
the $\{\ket{0}, \ket{1}\}$ basis, and quantum gates. However, other
methods may be possible. For example, one could engineer the
coupling between each ancilla block $a_x$ and its environment in such a
way that relaxation would place $a_x$ in the desired state
$\ket{0_E} + \ket{1_E}$.

Any method to realise a `factory'
which delivered reliable ancillas could be used to stabilize the
quantum computer. Suppose we have a factory which generates the required
ancilla states rapidly,
but imperfectly. In that case one could use a recursive purification
method to obtain ancillas of high fidelity, as shown in fig. 6.
This is closely related to the purification method
for entangled pairs of qubits which can make
quantum cryptography secure \cite{Deut96,Benn96}.

The main conclusions of this study are as follows. First, 
since quantum error correction is essentially a method in which
processing is used to fight noise, it is meaningless to specify
the noise level which can be tolerated without stating the
processing power required. The main parameters of the
processing power are the scale-up $N/K$, the slow-down $T/Q$,
and the degree of parallelism. It makes sense to distinguish
the noise level $\gamma$ per gate from the noise per time
step $\epsilon$ for resting qubits,
but it is important to include both contributions. Even
when $\epsilon$ is a hundred times smaller than
$\gamma$, memory errors can still be the dominant source of
noise.

In previous work I argued that the ancilla factory method is successful 
because it reduces the operations on the encoded qubits of the computer to 
a minimum. Here I have further shown that the ancilla state need only be
verified against a small subset of the errors it might have; this also
is important to the success of the method.

The analysis of the error rates and overheads shows that the ancilla 
factory is the best known method to stabilize quantum computation. 
Furthermore, block coding is better than concatenated coding using 
$7^L$-bit codes. The reason is that the efficiency of block codes more than 
compensates for the advantage of correcting low-order errors more often 
than high-order ones, which is the chief advantage of concatenated coding.
The best method may be a combined one, in which a more advanced block
code such as the Golay code is concatenated with a simple one
such as the 7-bit code.

In order to use these methods in a real quantum computer, it will
be necessary to achieve the correct noise properties, chiefly
the property of uncorrelated noise at different times and places.
A real computer might use small-scale methods applied to
individual physical qubits, combined with quantum coding on
a larger scale. The small-scale methods would depend on the
physics of the particular system, and their chief aim would
be to ensure that the noise left to be corrected through a QECC
would have the right properties.

I thank the Royal Society and St Edmund Hall, Oxford, for their
support.

\begin{figure}
\caption{Preparation of the state $\ket{0_E} + \ket{1_E}$
in the ancilla, including verification to elliminate sign errors in the final 
state (ie bit errors before the $H$ gates). Each horizontal line is a bare 
qubit. A circle at the left of a line represents preparation of $\ket{0}$. A 
small box represents measurement in the basis $\{ \ket{0}, 
\ket{1} \}$. If the measurement result is 1, the whole network is restarted. If 
it is zero the measured qubit is taken to be prepared in $\ket{0}$.} 
  \label{fig1}
\end{figure}

\begin{figure}
\caption{A simple equivalence used in the analysis. Each horizontal
line is an encoded qubit, ie one block. 
The network on the left is used in practice; the equivalent
network on the right helps to analyse the propagation of errors.}
\end{figure}

\begin{figure}
\caption{A complete correction of a single block. Each horizontal
line is a logical qubit, ie one block. A circle at the left of a
line represents preparation of $\ket{0_E}$ with verification
against bit errors (cf fig. 1). The $r$ syndromes are either
generated in parallel (a) or in series (b). The case $r=2$ is illustrated.}
\label{fig3}
\end{figure}

\begin{figure}
\caption{Failure probability, per correction, per block, 
(eq. (\protect\ref{P})) as
a function of gate error probability $\gamma$, for various codes.
By reducing $r$ (on average) it should be
possible to achieve the same $P$ with $\gamma$ increased
by a factor $\sim t$ (see text, section \protect\ref{s:d}), but
a more complete analysis is needed to confirm this. 
Dashed line: $[[7,1,3]]$; full line: $[[23,1,7]]$; dash-dotted:
$[[55,1,11]]$; dotted: $[[87,1,15]]$. (a) serial
repetition of syndromes ($s = s_s$), taking $n \epsilon = \gamma/2$.
(b) parallel repetition of syndromes ($s = s_p$), taking
$n \epsilon = 2 \gamma$.}
\label{fig4}
\end{figure} 

\begin{figure}
\caption{Reduced network for generation and verification of
the ancilla. This does the same job as fig. 1 but the verification
takes advantage of the structure of the generation network. No single
gate failure or memory error can produce undetected bit errors in
two qubits.}
\end{figure}

\begin{figure}
\caption{Recursive network for purification of ancillas. In this
illustration the ancillas are tested for both bit errors and
sign errors, and a single level of recursion is shown. 
Ancilla states which survive the purification can be fed
into the left of the network to obtain the next level.}
\end{figure}

\begin{table}
\begin{tabular}{clcccc}
       & units & conc.  & [[23,1,7]] & [[55,1,11]] & [[87,1,15]] \\
       &       & $7^3$ bit \\
$\gamma$  & $10^{-6}$ & 1  & 2.2  & 9   & 17    \\
$\epsilon$& $10^{-6}$ & 1  & 0.05 & 0.08  & 0.1   \\
$N$       & $10^5$    & 13 & 1.5  & 3.6  & 5.7   \\
$T$       & $10^{16}$ & 10 & 0.3  & 2    & 3.9   \\
parallelism & $10^4$  & 20 & 0.6  & 0.6  & 0.6
\end{tabular}
\caption{Error rates and overheads required to perform $Q = 2 \times 10^{10}$
computational steps on $K = 2150$ encoded qubits. The 
error rates for the $7^3$ bit concatenated code are taken from
a more careful analysis in \protect\cite{Pres97}. The other rates are
obtained from eq. (\protect\ref{P}), using
serial repetition of syndromes (2 parallel ancillas used $r$ times),
and putting $n \epsilon = \gamma / 2$ (cf fig. 4a).
Optimisation (see section \protect\ref{s:d}) would enable the results shown 
to be improved upon.} \end{table} 

\begin{table}
\begin{tabular}{clcccc}
       & units & conc. & [[23,1,7]] & [[55,1,11] & [[87,1,15]] \\
       &       & $7^3$ bit \\
$\gamma$  & $10^{-6}$ & 1  & 1.7  & 8   & 19    \\
$\epsilon$& $10^{-6}$ & 1 & 0.15 & 0.3  & 0.44   \\
$N$       & $10^5$    & 13 & 4.6   & 16   & 32    \\
$T$       & $10^{16}$ & 10 & 0.3  & 2    & 3.9   \\
parallelism & $10^4$  & 20 & 1.9  & 2.8  & 3.7
\end{tabular}
\caption{As table 1, but the results for block coding assume
each syndrome is extracted via $2r$ ancillas used in parallel,
and we take $n \epsilon = 2 \gamma$, (cf fig. 4b).}
\end{table}


\begin{thebibliography}{99}

\bibitem{CS} A. R. Calderbank and P. W. Shor, Phys. Rev. A {\bf 54}, 1098
(1996). 

\bibitem{Steane2} A. M. Steane, Proc. Roy. Soc. Lond. A {\bf 452}, 2551 (1996).

\bibitem{ft} P. W. Shor,
``Fault-tolerant quantum computation,''
in {\em Proc. 37th Symp. on Foundations of Computer 
Science}, to be published. 

\bibitem{Kita} A. Yu. Kitaev,
``Quantum error correction with imperfect gates,'' preprint.

\bibitem{concat} E. Knill, R. Laflamme, ``Concatenated quantum codes,''
(LANL eprint quant-ph/9608012)

\bibitem{Kitb} A. Yu. Kitaev, ``Quantum computing: algorithms and
error correction,''
Russian Mathematical Surveys, to appear.

\bibitem{KLZ96} E. Knill, R. Laflamme and W. H. Zurek,
``Accuracy threshold for quantum computation,''
(LANL eprint quant-ph/9610011)

\bibitem{AhBO96} D. Aharonov and M. Ben-Or,
``Fault-tolerant quantum computation with constant error,''
(LANL eprint quant-ph/9611025)

\bibitem{Got96} D. Gottesman, J. Evslin, S. Kakade and J. Preskill,
to be published (1996).

\bibitem{KLZ97} E. Knill, R. Laflamme and W. H. Zurek,
``Resilient quantum computation: Error Models and Thresholds,''
(LANL eprint quant-ph/9702058)

\bibitem{Pres97} J. Preskill,
``Reliable quantum computers,''
(LANL eprint quant-ph/9705031)

\bibitem{Steane97} A. M. Steane, Phys. Rev. Lett. {\bf 78}, 2252 (1997).

\bibitem{Gott} D. Gottesman, Phys. Rev. A {\bf 54}, 1862 (1996).

\bibitem{CRSS} A. R. Calderbank, E. M. Rains, P. W. Shor and N. J. A. Sloane,
Phys. Rev. Lett. {\bf 78}, 405 (1997).

\bibitem{DiVS} D. P. DiVincenzo and P. W. Shor, Phys. Rev. Lett. {\bf 77},
3260 (1996).

\bibitem{Got97} D. Gottesman,
``A theory of fault-tolerant quantum computation,''
(LANL eprint quant-ph/9702029)

\bibitem{Knill} E. Knill and R. Laflamme, Phys. Rev. A, to be published.
(LANL eprint quant-ph/9604034).

\bibitem{Benn} C. H. Bennett, D. DiVincenzo, J. A. Smolin and W. K. Wooters, 
Phys. Rev. A, submitted (LANL eprint quant-ph/9604024). 

\bibitem{Steane1} A. M. Steane, Phys. Rev. Lett. {\bf 77}, 793 (1996).

\bibitem{SQECC} A. M. Steane, Phys. Rev. A {\bf 54}, 4741 (1996).

\bibitem{Golay} M. J. E. Golay, Proc. IEEE {\bf 37}, 657 (1949).

\bibitem{MacWS} F. J. MacWilliams and N. J. A. Sloane,
``The theory of error-correcting codes,''
(North-Holland, Amsterdam, ninth impression 1996.)

\bibitem{Beck} D. Beckman, A. Chari, S. Devabhaktuni and J. Preskill,
Phys. Rev. A {\bf 54}, 1034 (1996).

\bibitem{Zalka} C. Zalka,
``Threshold estimate for fault tolerant quantum computing,''
(LANL eprint quant-ph/9612028)

\bibitem{Deut96}
D. Deutsch, A. Ekert, R. Jozsa, C. Macchiavello, S. Popescu
and A. Sanpera, Phys. Rev. Lett. {\bf 77}, 2818 (1996).

\bibitem{Benn96} 
C. H. Bennett, G. Brassard, S. Popescu, B. Schumacher,
J. A. Smolin and W. Wootters, Phys. Rev. Lett. {\bf 76}, 722 (1996).


\end{thebibliography}
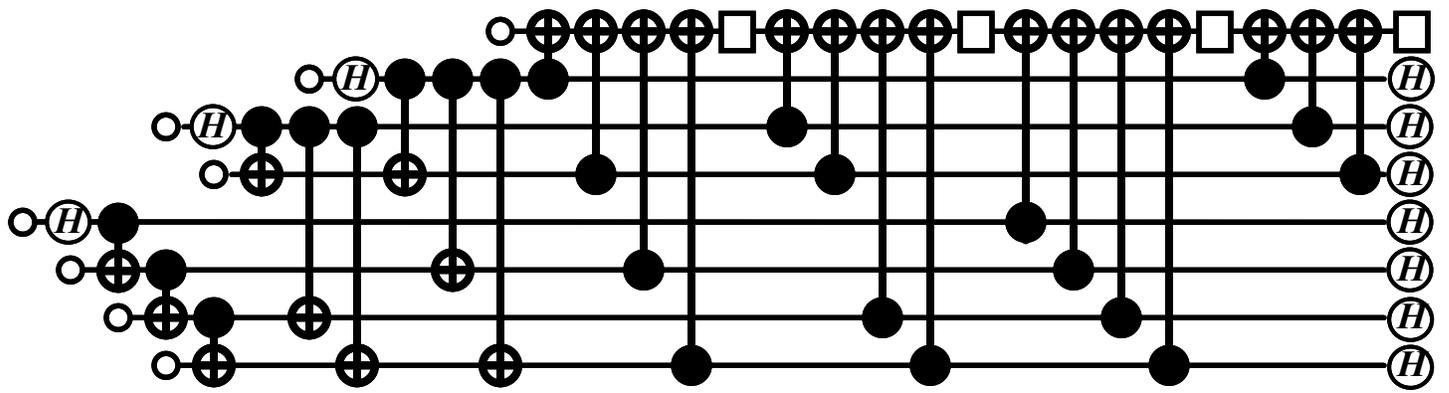
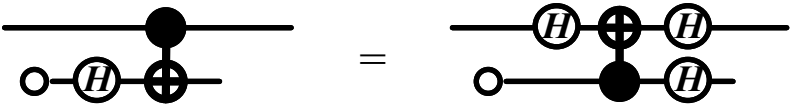
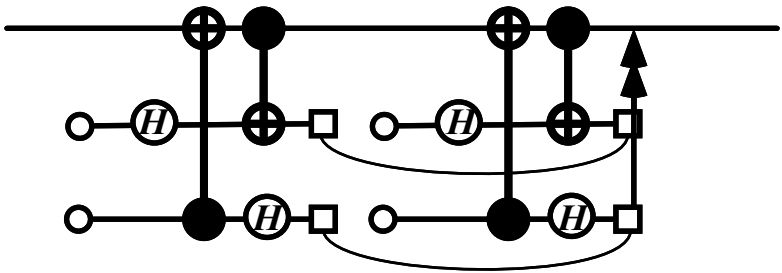
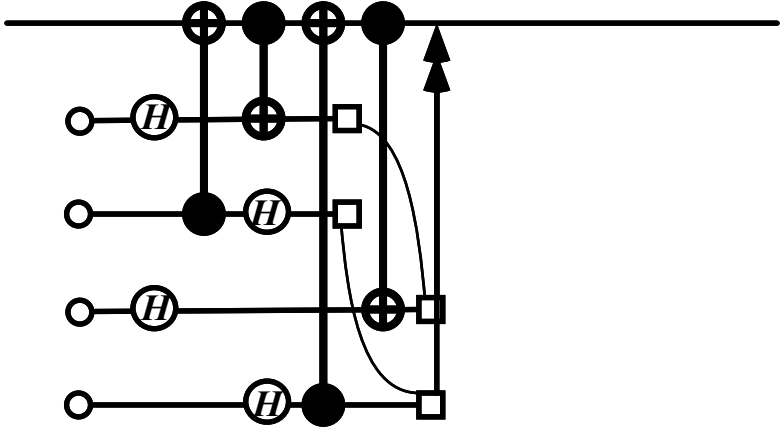
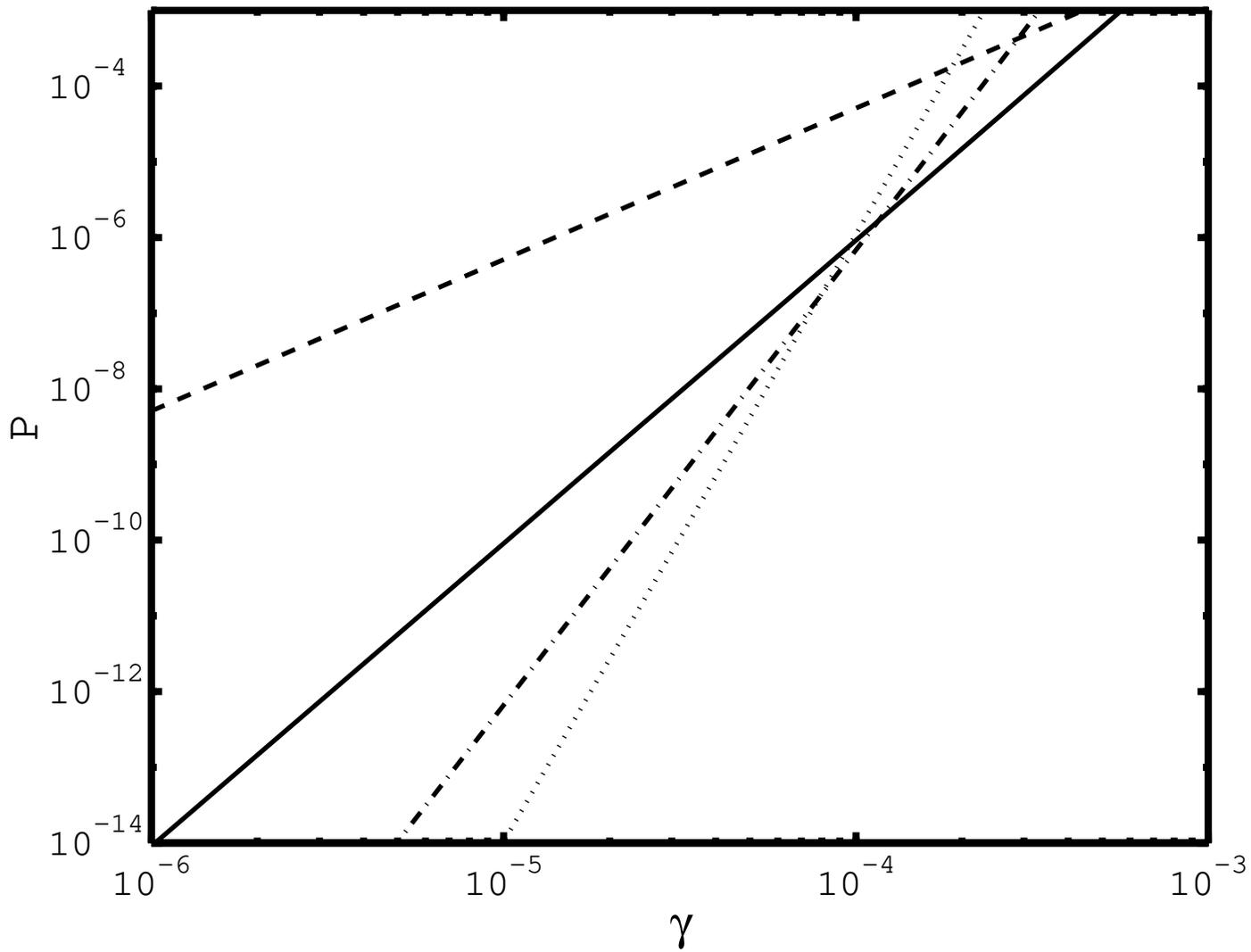
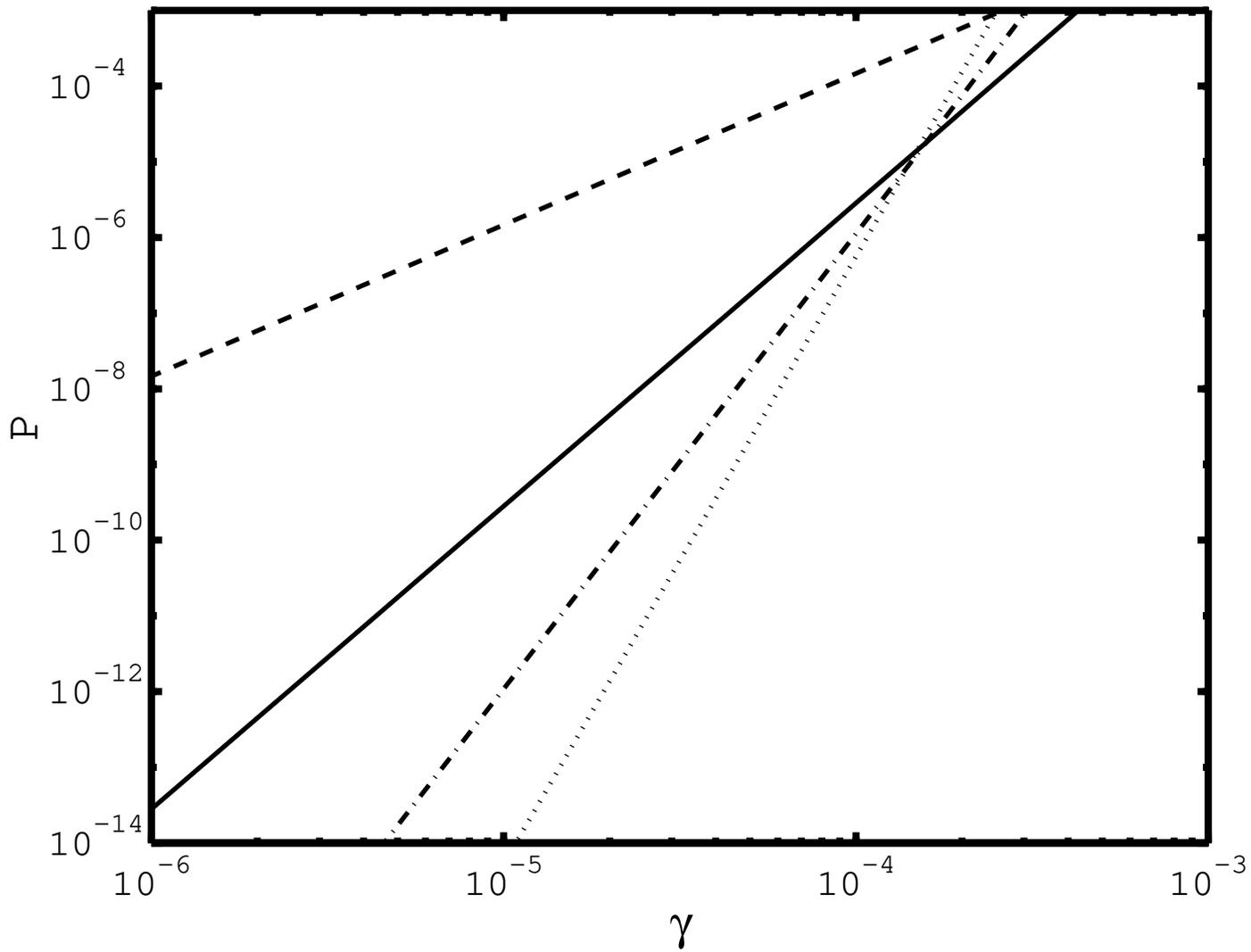
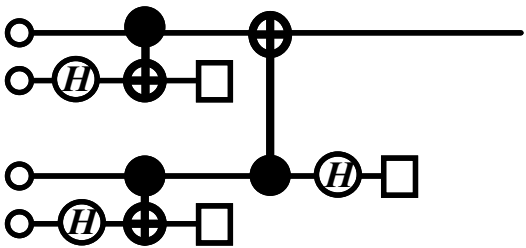
\end{document}